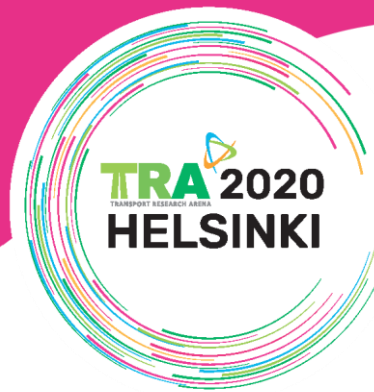

# Passive Wi-Fi Monitoring in Public Transport: A case study in Madeira Island

1. Miguel Ribeiro[a], Bernardo Galvão[b], Catia Prandi[c], Nuno Nunes[a]

[a]*ITI/LARSYS, Instituto Superior Tecnico - ULisboa, Lisboa, Portugal*
[b]*ITI/LARSYS, Funchal, Portugal*
[c]*University of Bologna, Bologna, Italy*

Abstract

Transportation has become of evermore importance in the last years, affecting people's satisfaction and significantly impacting their quality of life. In this paper we present a low-cost infrastructure to collect passive Wi-Fi probes with the aim of monitoring, optimizing and personalizing public transport, towards a more sustainable mobility.
We developed an embedded system deployed in 19 public transportation vehicles using passive Wi-Fi data. This data is analyzed on a per-vehicle and per-stop basis and compared against ground truth data (ticketing), while also using a method of estimating passenger exits, detecting peak loads on vehicles, and origin destination habits.
As such, we argue that this data enables route optimization and provides local authorities and tourism boards with a tool to monitor and optimize the management of routes and transportation, identify and prevent accessibility issues, with the aim of improving the services offered to citizens and tourists, towards a more sustainable mobility.

## 2. Introduction

According to the World Commission on Environment and Development (1987), already since the last century, the world we live was becoming ever more data-driven. Urban environments were and continue to become increasingly more instrumented and interconnected as pervasive sensors and devices are generating vast amounts of data, which in turns produces new insights and knowledge about the world we live in. Objects' trajectories recorded by GPS enabled devices, Steg and Gifford (2005) and European Environment Agency (2018), and network-based tracking technologies recorded by wireless networks as done by Høyer (2000), are all generating data, enabling new interpretations and insights useful to society at large.

These increasingly accurate spatiotemporal data can be leveraged to understand better phenomena ranging from epidemics, ecology, global warming, urban dynamics, and as shown by Verbeek and Mommaas (2008) to other social issues such as crime, migrations, traffic and land use. One of the main interest areas is tourism, a continuously growing and important sector for many regions and countries worldwide for which it is a central and vital source of welfare. Goodchild (2007) were already starting to use spatiotemporal data to understand the evolution of tourism as a social commodity. Goodchild (2007) has also made advances in transportation and communication technologies impacting aspects such as the choice and spatial behavior of destinations.

In public transportation, it is pivotal for the companies to build an understanding of their users, about how they use the transports, where they enter and leave the bus. This enables the calculation of important measures such as the times of higher occupancy, waiting times, among other statistics that can be used to optimize bus routes and scheduling, important issues to consider in sustainable mobility. In this case study, we want to detail how our low-cost solution can be used to support transport companies in planning decisions about how to schedule buses and where to locate the bus stops, taking advantaged of both real time data and historical data. In fact, considering the latter one, analyzing historical data related to counting of people on the bus, the mobility manager can see in which hours a particular bus becomes too crowded and decide if intensification the service offer for that specific time slot so to better meet the customer needs, and maybe, suspend some bus rides if the number of people using that service is too low to explain that cost in term of $CO_2$ consumption improving the sustainability.

The focus of this research revolves around the use of low-cost Wi-Fi tracking and data mining technologies as an alternative to existing methods of studying mobility patterns at scale. The system was installed in 19 buses of a stakeholder bus company to quantify and generate useful data about the presence and movement of people around the main city of a popular touristic Island destination. To assess the effectiveness of the method we compared the results with the ticketing system of the bus company stakeholders to gather data about events and flows of people in the main gateways such as airports and ports.

## 3. Related Work

People counting has always been a useful tool to control environments. Several challenges are faced depending on the type of location, how constrained or how intrusive the system is, and how costly it is to install. Several attempts have looked at Wi-Fi as a means to infer crowd densities. Existing commercial systems are expensive and proprietary, enabling access only to large corporations and telecom providers, which own the software and communication infrastructures but are entirely detached from the underlying realities and opportunities.

Several studies have attempted to locate or count the number of people in specific locations. Most of them use wireless technologies, with a large part of them exploring technologies such as Dixon et al. (2013) with RFIDs or such as Baniukevic, Jensen, and Lu (2013) with Bluetooth. Bonné et al. (2013); Meneses and Moreira (2012) and Song and Wynter (2017) used Wi-Fi technology to capture human mobility information in highly crowded areas such as football games, universities campuses and hospitals. The motivations of many of these studies are diverse, Bonné et al. (2013) looks at energy waste on scanning methods, Meneses and Moreira (2012) focus on realistic facility management and planning, while others, such as Kjærgaard et al. (2012) and Sapiezynski et al. (2015) looked at crowding factors, flock detection and waiting times, speed and frequent paths and even social information like popularity of events (in the case of Song and Wynter (2017) singers in concerts).
Several networking infrastructure vendors offer geo-marketing solutions for organizations deploying large Wi-Fi networks (such as shopping malls, hotels, and airports). Driven by the explosion of digital data the possibilities of understanding the dynamical and topological stability of large networks is increasing. Barabasi and Albert (1999) showed that the development of large networks is governed by robust self-organizing phenomena that go beyond the particulars of the individual systems. Exploring several datasets describing the topology of large networks the authors showed that large networks self-organize into a scale-free state, a feature unexpected by all existing random network models.

There are several techniques to analyze and synthetize mobility information from tracking data. For instance, early work proposed by Zheng et al. (2009) used tree-based hierarchical graphs from GPS trajectories to mine exciting locations and classical travel sequences. Their work was based on modeling GPS logs into trajectories (sequence of GPS logs based on their time series) and stays (a geographic region where a GPS log is observed over a period of time). Besides, user profiling techniques were also used in social platforms, such as Twitter and Foursquare both presented by Cheng et al. (2011) and Chong, Dai, and Lim (2015), where the users were classified according to the trip durations, the distance between locations, and past locations and searches to predict their travel paths.

Focusing on public transportation, Farshad, Marina, and Garcia (2014) analyzed the presence Wi-Fi usage on a set of Wi-Fi access points by traveling by bus in the city of Edinburgh. The authors focused on categorizing locations by cafes, shopping areas, houses and apartments, trying to characterize the channels and frequency used, and mapped the city finding a prominence in 2.4 GHz, and channels 1, 6 and 11 being the most common.

Song and Wynter (2017) tried to determine the timetables using public transport monitoring through passive Wi-Fi. By using the captured data, they analyzed the ratio of number of stations where a device is recorded, the dwell time and headway as the time difference before the departure of this train and the arrival of next train. They tried to cluster similar behaviors between users that entry and exit in similar bus stops, however these models were susceptible to a flaw that it is prone to "inventing" trains by creating clusters that do not correspond to any physical train. From the reported data, only one of the clusters of hit rate of the baseline achieved a spectral clustering higher than 0.55.

Baeta, Fernandes, and Ferreira (2016) did a study using a Cisco Access point on one vehicle in one bus route. The data from the bus was crossed with geolocation data from a GPS data and each record included mainly the bus id, route id, router way, trip number and timestamp so they could be referenced after being sent to the cloud server database. Also, to reduce the amount of redundant information, the Wi-Fi log file had several entries per second for each client for a single scanning, which were grouped into a single record. The Wi-Fi data was matched with the GPS data via timestamp, and then matched with the vehicle and route scheduling. The passenger counting estimation was done by 4 arbitrary rules: 1) sampled by segments minus a time-delta of 10 s, 2) sampled by segments until reaching a stop zone of 100 m radius; 3) passenger valid if present on 2 consecutive segments and 4) passenger valid if present on 3 consecutive segments. The results show mixed outcomes to inferring the precision of real count vs detected count, ranging from 0 to 100 % depending on the route stop of the analyzed route.

Most of the analysis reviewed above focus on either small supervised tests in localized campuses, or larger unsupervised tests in small cities and on moving vehicles with limited routes and buses. While campuses and research rooms offices provide good test beds with ground truth, they have a common biased user base and typology. Our study focuses on the monitoring of several public transport vehicles over several routes, analyzing the data modeling to infer vehicle load, and origin destination matrices.

## 4. The Deployment

Our system was deployed into bus vehicles with a maximum load of 83 passengers and built exploiting existing inexpensive commercial Wi-Fi routers deployed in 19 buses The routers were flashed with an open-source GNU/Linux based firmware (openWRT) to run a program for embedded devices. The routers operate in monitoring mode and the probe request information is stored in a central MySQL database after the vehicles reach the base station. The MAC addresses detected in the probe requests were locally transformed using a cryptographic hash function to prevent access to the original identifiers that could be used to compromise the privacy of users. There is no direct link between the MAC address and the user, however, it is possible to extract information regarding the individual's habits and potential social connections. The server side components perform the calculations and optimizations required for analyzing the captured data and provide the results through a web server to the clients. When at the base station the Wi-Fi routers are connected to a VPN allocated on the server to allow their remote management, and the scripts processing the data interact with several external services and APIs. Then the data from the vehicle, route scheduling and ticketing information is provided to perform our analysis.

In order to collect the mobility data, we designed and developed an infrastructure exploiting: (a) a passive Wi-Fi tracking system, named Beanstalk (Nunes et al. (2017)), that passively collects Wi-Fi requests from devices that people carry in their daily commutes (e.g. smartphones) integrated in the Beanstalk infrastructure, to gather data about transport systems occupancy and movements. Moreover, we installed our sensory infrastructure in 19 buses, in order collect data related to public transport usage in the whole city. The overall system architecture is shown in Figure 1 and represents the Wi-Fi sensor, the server, the database, the bus terminal and the communications between them. It also displays the possibility to connect several USB sensors to the nodes (Wi-Fi routers).

In the next subsections, we provide details about the Wi-Fi tracking used to collect geolocated data, and the deployment of the system infrastructure in Madeira island, exploiting the data collection data from the bus routes.

**5. Methods**

The deployment was conducted in Madeira, a Portuguese archipelago in the north Atlantic, where tourism is an important sector of the region's economy and constitutes an important market for local products, with a relevant impact in the island mobility. The routers were installed in 19 vehicles of the main city bus company. These vehicles were scheduled for different routes, meaning that we did not have control, a priori, of the routes that the data was referring to. We also had access to the bus scheduling and the ticketing validations for a period of 2 and a half months. A ticketing dataset is used as ground truth data for the passengers that entered the vehicles, containing a timestamp; bus route; and the bus stop associated with each ticket validation. Sensitive passenger information is omitted.

The online system provided web-based origin-destination matrices to the bus company as well as the daily and hourly unique detections of devices, which provided information about vehicle load. Both the data from which estimates are made, and the target ground truth data have to be mapped to bus trips and bus stops with regards to the time, the route and direction of the trip.

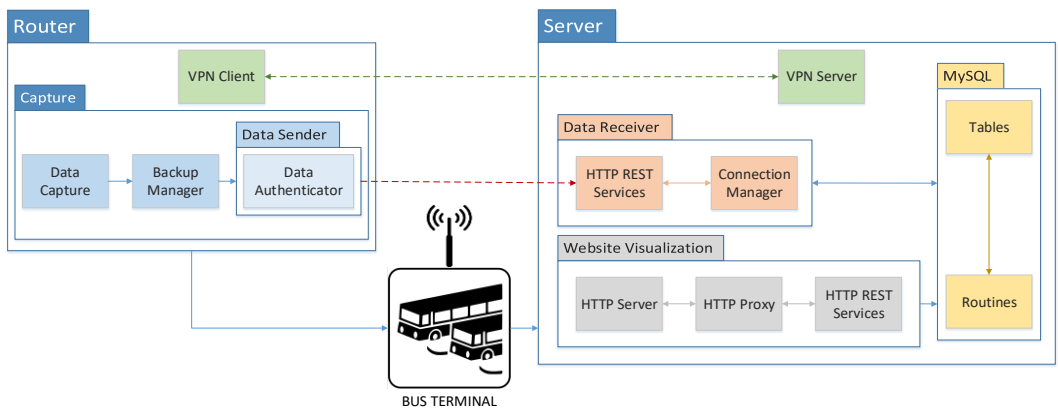

Figure 1 - Sensor system components and server

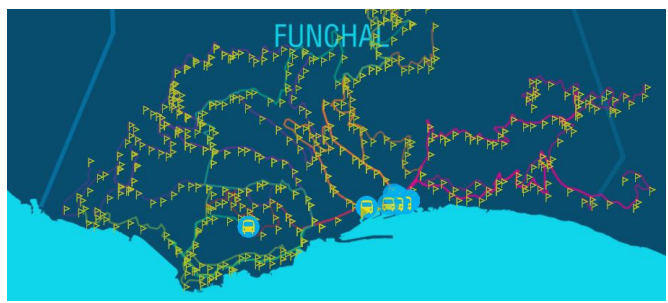

Figure 2 - Bus routes and stops analyzed in the city of Funchal

One aspect to consider is the radius of detection of networked devices. For instance, it is not desirable to detect devices that are outside the bus; likewise, if the detection radius is rather short, the risk of not detecting all the relevant devices within the bus becomes present. In order to address this aspect, a set of RSSI (Received Signal Strength Indicator), as proxies to physical distances is considered, as the RSSI measures the power level that a radio client device is receiving from a router. For this test, the RSSI signal set is composed by -128 dB (very weak signal, corresponding to greater distance) and the range from -85 dB (moderately weak) to -55 dB (strong signal, corresponding to a shorter distance) in steps of +5 dB. The most appropriate signal to count passengers is determined in the results section.

*5.1. Mapping to bus stops*

This subsection addresses how Wi-Fi probe requests are mapped to the bus stops in terms of origin-destination. In general, this mapping follows the rationale illustrated by Figure 3.
- a given device is assumed to have entered a bus at point $t$ if its detection occurred between time t and $t + 1$;
- and is assumed to have left the bus at time $t$ if its last detection occurred after $t$.

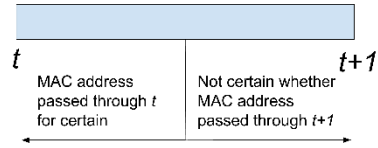

Figure 3 - Logic behind mapping MAC addresses to bus stops based on the timing of their probe requests. On *t*, it is considered that networked devices may take time to emit a probe request after entrance in the bus vehicle. On *t+1*, it is considered that a MAC address captured during this period may not have resumed its bus trip beyond stop t+1.

The number of passengers between $t$ and $t + 1$ can be identified by the number of different MAC addresses detected within that period (denoted by $w$); or deduced from Wi-Fi detection of entries and exits as described and following the sequential logic that the current number of passengers (denoted by $c$) at the starting point of a bus trip is simply given by the detected entrances (denoted by $i$):

$$c_0 = i_0 \tag{1}$$

And at the following bus stops, the current number of passengers is given by:

$$c_t = c_{t-1} + i_t - o_t \tag{2}$$

Where $i_t$ stands for the number of passengers that got in a bus at stop $t$ (i.e. the number of MAC addresses first detected at stop $t$) and $o_t$ stands for the number of passengers that got out of a bus at stop $t$ (i.e. the number of MAC addresses last detected after stop $t - 1$). Ideally, the relation $w_t = c_t$ ought to hold true and consistent and is tested in the Results section.

*5.2. MAC address randomization*

In spite of the ubiquity of networked handheld devices, recent privacy concerns on the tracking of people in the EU have motivated a policy of MAC address randomization, the implementation of which is a decision of the vendor. For instance, Microsoft makes Windows 10 randomize a MAC address at every probe request upon user choice; Android generally randomizes a MAC address per SSID (Service Set Identifier, a network identifier), but MAC address randomization greatly depends on the implementation from smartphone manufacturers. In order to account for this issue, it is assumed that MAC addresses that are detected only once in a bus trip are randomized; all the remaining MAC addresses that are detected more than once are considered to be non-randomized.

**6. Results**

Denoting ground truth ticket validation data as $b_t$, it is desirable that the following relationships hold:
- (A) $b_t = i_t$, meaning that the number of entries on a bus estimated by Wi-Fi ($i_t$) - where an entry is estimated by the first time a MAC address is detected on a bus trip - ought to be equal to the ground truth number of ticket validations ($b_t$) at time t.
- (B) $w_t = c_t$ which is the same as asking the question: is the number of different MAC addresses detected between $t - 1$ and $t$ the same as the number of passengers deduced from Wi-Fi estimated entries and exits, i.e.

$c_t = c_{t-1} + i_t - o_t$. This represents a concern of data consistency from the Wi-Fi sensors so as to consider if the frequency of probe request interception is high enough, or, alternatively, to consider if devices emit probe requests at a sufficient rate.

However, before evaluating these, it is necessary to determine the optimal parameters for minimum RSSI signal; and whether to include random MAC addresses (i.e. those that appear only once in the log of a bus trip). For these purposes, 5 different metrics were calculated per bus trip, these being:

1. Mean Absolute Error (MAE)
2. Median Absolute Error
3. Standard Deviation of Absolute Error
4. Root Mean Square (RMS) Error
5. R2 Score

The reason for including more than one metric lies behind the idea that the perfect parameter choice of minimum RSSI and inclusion of random MAC addresses should be one that performs well across all of these metrics. Thus, each of these metrics was calculated for every combination of case ((A) and (B)); and minimum RSSI; and bus trip (naturally, an instance of a bus route). This computed data aided the choice of optimal minimum RSSI to pick per route and whether to use random MAC addresses.

*6.1. Optimal minimum RSSI*

Table 1 shows the modal optimal minimum RSSI across all cases and metrics, which is picked for each route - i.e. the minimum RSSI value that has the largest number (i.e. share) of combinatorial instances (of metric and case (A) or (B) of every bus trip belonging to a route) for which it performed better. E.g.: for route #1, a minimum RSSI of -55 performed better in 83% of the instances of every case, metric and bus trip belonging to that route.

This is a way to satisfy multiple objectives per route. For the majority of the routes, the best minimum RSSI is -55, which, at first sight, tells us Wi-Fi estimations approximate better to the ticket validations with the shortest radius available in our data.

Table 1 - Share of modal best min RSSI across all metrics and cases per route. A minimum RSSI of -55 performs generally better.

| Route Number | Min RSSI | Share | Route Number | Min RSSI | Share |
|---|---|---|---|---|---|
| 1 | -55 | 0.83 | 36 | -55 | 0.57 |
| 10 | -75 | 0.53 | 36A | -80 | 0.40 |
| 12 | -55 | 0.77 | 38 | -55 | 0.63 |
| 13 | -55 | 0.93 | 4 | -55 | 0.93 |
| 16 | -55 | 0.63 | 44 | -55 | 0.90 |
| 2 | -55 | 0.80 | 45 | -55 | 0.83 |
| 20 | -55 | 0.97 | 47 | -55 | 0.77 |
| 21 | -55 | 0.93 | 48 | -55 | 0.83 |
| 24 | -55 | 0.63 | 49 | -55 | 0.93 |
| 3 | -55 | 0.53 | | | |

*6.2. Random vs non-random MAC addresses*

For each of the cases (A) and (B), the Wilcoxon signed-rank test was computed on a route-by-route basis, testing whether the distribution of RMSE results including random MAC addresses differs with statistical significance of the distribution of RMSE results that exclude random MAC addresses. The results are summarized in Table 2 and Table 3 (only the routes with more than 30 trips available in the data are included so as to provide a statistically significant test). As it can be observed, the results of the measurements without random MAC addresses (denoted by the suffix "_ns" in the Tables), perform significantly better than their random-inclusive counterparts. As such, it is considered from this point onward that the optimal choice is to exclude potential random MAC addresses.

*6.3. Performance*

Having fixed minimum RSSI to -55 and decided to consider only non-random MAC addresses, it is finally possible to discuss the performance on ground truth data. Table 4 aggregates the mean results with these parameter choices per route (again, including only those with more than 30 trip samples) and per case for three metrics: MAE for interpretability; RMSE; and R2 score to consider how much of the variance of the target variable is explained by the proposed inputs for each case. Looking at MAE, the results look encouraging: in general, the mean absolute error is not greater than 1. The confidence intervals on confirm this. However, taking R2 scores into account it is concluded that our method of estimating passenger entries performs worse than the mean of the target variable (number of ticket validations in case (A)).

From the vehicle load, the origin destination matrixes were also calculated. An example is shown in Table 6.

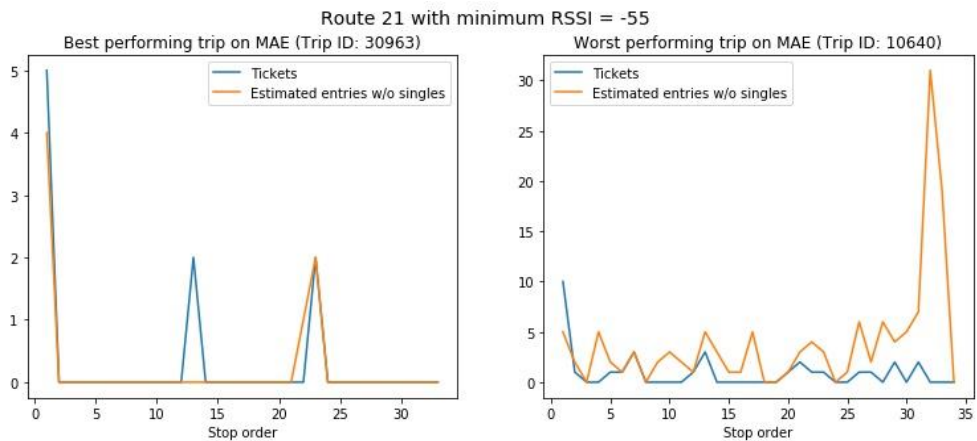

Figure 4 –Wi-Fi detected devices vs ticketing validation count over bus stops of route 21.

Figure 4 shows the number of passenger entries is represented by the Y axis, where the blue line represents ground truth data and the yellow line represents entries estimated from Wi-Fi probe requests; the X axis represents the bus stops in a full trip of the vehicle on the route 21. On the left is a representative case of a trip performing well on Mean Absolute Error (MAE), which happens when there are almost no passenger entries. On the right, a representative case of a trip where our method of counting entries does not perform so well on MAE, but is able to follow some peaks of the ground truth data.

Table 2 - RMSE minimizing case and Wilcoxon sum-rank p-value for case (A), proving that keeping only non-single (ns) MAC addresses is better.

| Route Number | Case | Wilcox pval |
| --- | --- | --- |
| 1 | A_ns | 8.189422e-08 |
| 12 | A_ns | 8.189422e-08 |
| 13 | A_ns | 8.189422e-08 |
| 2 | A_ns | 8.189422e-08 |
| 21 | A_ns | 8.189422e-08 |
| 4 | A_ns | 8.189422e-08 |
| 45 | A_ns | 8.189422e-08 |
| 47 | A_ns | 8.189422e-08 |
| 48 | A_ns | 8.189422e-08 |

Table 3 - RMSE minimizing case and Wilcoxon sum-rank p-value for case (B), proving that keeping only non-single (ns) MAC addresses is better.

| Route Number | Case | Wilcox pval |
| --- | --- | --- |
| 1 | B_ns | 2.313602e-07 |
| 12 | B_ns | 2.313602e-07 |
| 13 | B_ns | 2.313602e-07 |
| 2 | B_ns | 2.313602e-07 |
| 21 | B_ns | 2.313602e-07 |
| 4 | B_ns | 2.313602e-07 |
| 45 | B_ns | 2.313602e-07 |
| 47 | B_ns | 2.313602e-07 |
| 48 | B_ns | 2.313602e-07 |

Table 4 - 95% confidence intervals, mean and standard deviation of R2 scores per route and case (n>=30 trips).

| Route | Case | Lower ci | Upper ci | Count | Mean | Std |
|---|---|---|---|---|---|---|
| 1 | A_ns | -0.189 | -0.113 | 133 | -0.151 | 0.223 |
|   | B_ns | 0.165 | 0.318 |  | 0.242 | 0.447 |
| 12 | A_ns | -0.437 | -0.118 | 58 | -0.278 | 0.606 |
|    | B_ns | -0.957 | -0.390 |  | -0.673 | 1.078 |
| 2 | A_ns | -0.192 | -0.036 | 74 | -0.114 | 0.335 |
|   | B_ns | 0.094 | 0.324 |  | 0.209 | 0.495 |
| 45 | A_ns | -0.143 | 0.047 | 64 | -0.048 | 0.380 |
|    | B_ns | -0.939 | -0.188 |  | -0.563 | 1.503 |
| 47 | A_ns | -0.362 | -0.065 | 54 | -0.214 | 0.543 |
|    | B_ns | -1.090 | -0.450 |  | -0.770 | 1.172 |
| 48 | A_ns | -0.200 | -0.095 | 62 | -0.148 | 0.205 |
|    | B_ns | -0.560 | -0.091 |  | -0.325 | 0.923 |

Table 5 - Mean results on cases (A) and (B) per route (n>=30 trips).

| Route | Case | MAE | RMSE | R2 score |
|---|---|---|---|---|
| 1 | A_ns | 0.858 | 2.199 | -0.151 |
|   | B_ns | 0.206 | 0.432 | 0.242 |
| 12 | A_ns | 0.763 | 1.639 | -0.278 |
|    | B_ns | 0.683 | 0.933 | -0.673 |
| 2 | A_ns | 1.219 | 2.619 | -0.114 |
|   | B_ns | 0.470 | 0.837 | 0.209 |
| 45 | A_ns | 0.859 | 1.676 | -0.048 |
|    | B_ns | 0.580 | 0.819 | -0.563 |
| 47 | A_ns | 0.466 | 1.688 | -0.214 |
|    | B_ns | 0.653 | 0.864 | -0.770 |
| 48 | A_ns | 0.599 | 1.289 | -0.148 |
|    | B_ns | 0.533 | 0.818 | -0.325 |

Table 6 - Origin/Destination matrix for route 8A with a total of 48 detections.

| Stop | 10 | 13A | 31 | 33 | 35 | 46 | 48 | 650 | 652 | 700 | 742 | 763 | 1020 | 1357 | 1377 |
|---|---|---|---|---|---|---|---|---|---|---|---|---|---|---|---|
| 2 | 0 | 0 | 0 | 0 | 0 | 0 | 0 | 0 | 0 | 0 | 0 | 0 | 0 | 0 | 0 |
| 10 | 0 | 0 | 0 | 0 | 0 | 0 | 0 | 0 | 0 | 0 | 0 | 0 | 0 | 0 | 0 |
| 13A | 2 | 0 | 0 | 6 | 0 | 1 | 0 | 0 | 0 | 0 | 0 | 0 | 0 | 0 | 0 |
| 31 | 4 | 0 | 0 | 0 | 6 | 10 | 0 | 0 | 0 | 1 | 0 | 0 | 0 | 0 | 6 |
| 33 | 0 | 0 | 0 | 0 | 0 | 0 | 0 | 0 | 0 | 0 | 0 | 0 | 0 | 0 | 0 |
| 35 | 0 | 0 | 0 | 0 | 0 | 0 | 0 | 0 | 0 | 0 | 0 | 0 | 0 | 0 | 0 |
| 46 | 1 | 0 | 0 | 0 | 0 | 0 | 0 | 0 | 0 | 0 | 0 | 0 | 0 | 0 | 0 |
| 48 | 0 | 1 | 0 | 0 | 0 | 0 | 1 | 0 | 0 | 0 | 0 | 0 | 0 | 0 | 0 |
| 650 | 0 | 0 | 0 | 0 | 0 | 0 | 0 | 0 | 0 | 1 | 0 | 0 | 0 | 0 | 0 |
| 652 | 0 | 0 | 0 | 2 | 0 | 0 | 0 | 0 | 0 | 0 | 4 | 0 | 0 | 0 | 0 |
| 700 | 0 | 0 | 0 | 0 | 0 | 0 | 0 | 0 | 0 | 0 | 0 | 0 | 0 | 0 | 0 |
| 742 | 0 | 0 | 0 | 0 | 0 | 0 | 0 | 0 | 0 | 0 | 0 | 0 | 0 | 0 | 0 |
| 763 | 0 | 0 | 0 | 0 | 0 | 0 | 0 | 0 | 0 | 0 | 0 | 0 | 0 | 0 | 0 |
| 1020 | 0 | 0 | 0 | 0 | 0 | 0 | 0 | 0 | 0 | 0 | 1 | 0 | 0 | 0 | 0 |
| 1357 | 0 | 0 | 0 | 0 | 0 | 0 | 0 | 0 | 0 | 0 | 0 | 1 | 0 | 0 | 0 |

## 7. Discussion

In the previous section, we detail the methods and results obtained from monitoring passengers in public transport vehicles with passive Wi-Fi monitoring techniques. In particular, they cover three main issues of interest, providing: (i) a model to infer the number of people from the Wi-Fi detections, tested against ground truth data for different distances captured; (ii) vehicle load to increase the monitorization and the understanding of how citizens use public transport; (iii) origin/destination matrices to understand where the passengers exit the vehicles. More generally, other scenarios can be considered, involving all the different aspects of the mobility and the specific interests of local communities who are directly affected by the mobility flow of tourists and residents. In fact, using the collected data, it is possible to obtain the citizens' most common mobility patterns and, accordingly, activate strategies to expand (or break) these patterns including less known but more unique and authentic places. This can become really feasible thanks to our innovative solution, that can stimulate the awareness in sustainable mobility for a sustainable tourism, and represent a virtual place where to meet the different communities' needs. Regarding the data processing, following the considerations of the results, the best parameters of minimum RSSI and whether to keep single MAC addresses were attempted to be found. In general, optimizing for several metrics, it was found that a strong RSSI signal and filtering out single MAC addresses was preferable (note that, according to how Wi-Fi counts were being mapped to bus stops, these single-appearance MAC addresses would be entering at a bus stop and exiting at a next one).

Finally, we measured how the Wi-Fi counting estimations fared against truth data. The confidence intervals on results of R2 scores, found that, for the most part, these scores are negative - especially when taking into account that route 1 is an important route and very utilized in a high degree by both locals and tourists alike. It is therefore concluded that Wi-Fi estimation of passengers, at least with the data processing proposed in this paper, is not a suitable solution to estimate entries and, as consequence, exits. It is worth to notice, that the community's authorities remain in charge to implement sustainable strategies and take decisions, but the tool can provide an easy way to visualize and detect relevant/unusual situations and to enhance awareness, supporting and facilitating the communication between the different stakeholders' needs. In fact, a sustainable development requires joining the effort of the different communities and authorities, in order to consider its social dimension and to become an effective framework to improve sustainable mobility and tourism.

## 8. Conclusion and Future work

In this paper, we present a low-cost infrastructure capable of collecting location-based data and to provide services to the different communities involved with the aim of soliciting the adoption of optimization of routes and vehicle scheduling by observing real time and historical data. The infrastructure exploits a Wi-Fi passive tracking system deployed the infrastructure in 19 vehicles of public transport. We propose a data processing pipeline, starting with the features of the bus trip logs, in which it is described how ground truth data is mapped to bus stops and the challenges that come from relying on these logs. Wi-Fi data can be processed with regards to its RSSI with a way to exclude potentially random MAC addresses by filtering out the ones that show up only once in a bus trip.

Here we consider a real application to show how the collected data can be put to good use to improve the daily mobility experience, employing sustainable mobility. Moreover, we argue that our solution represents an opportunity to investigate the possibility to meet the different needs. As a general final aim of the infrastructure, this could provide mobility-related information not only to the route planning entities, but also empowering the different communities of stakeholders by collaborating in the sustainable development of transportation.

Turning data into information is ever more relevant nowadays. While our findings revealed that the proposed system has some drawbacks due to the passive nature of probing adopted, which are very dependent to the terminal Wi-Fi probe frequency, a natural extension of this work is also for the monitoring of crowd-related public transport service levels and disruptions, including alerts, waiting line detections at kiosk. In the future, we also plan to evolve this platform into a fully community network capable of offering other personalized services to communities, exploring crowdsourcing and crowdsensing. Nonetheless, we achieved the central goal to provide realistic information that reflects realistically the behavior of public transport users who make use of the vehicles and services offered. Thus, the proposed methods can provide valuable sources of information, e.g. regarding building, path and service utilization, for supporting transport planning activities.

## 9. Declarations

*9.1. Availability of data and material*

The datasets used and/or analyzed during the current study are available from the corresponding author on reasonable request.

*9.2. Competing interests*

The authors declare that they have no competing interests

*9.3. Funding*

This work was done under the CiViTAS Destinations project measure MAD2.2, with the support from the FCT grant SFRH/BD/135854/2018 and ARDITl-CIVITAS-DESTINATIONS/2018/002.

*9.4. Authors' contributions*

The first and second authors made substantial contributions to the data processing and all authors have drafted the work and substantively revised it.

*9.5. Acknowledgements*

This work had the support of the public transport company Horários do Funchal under CiViTAS Destinations measure MAD2.2.